\documentclass[aps,prl,twocolumn,superscriptaddress,showpacs, floatfix, nofootinbib]{revtex4}
\usepackage{comment}
\usepackage{graphicx}
\usepackage[dvips]{color} 
\usepackage{bm}

\newcommand{\met}{\ensuremath{\not \!\! E_T}}

\begin{document}

\title{Search for Technicolor Particles Produced in Association with a $W$ Boson at CDF}

\affiliation{Institute of Physics, Academia Sinica, Taipei, Taiwan 11529, Republic of China} 
\affiliation{Argonne National Laboratory, Argonne, Illinois 60439} 
\affiliation{University of Athens, 157 71 Athens, Greece} 
\affiliation{Institut de Fisica d'Altes Energies, Universitat Autonoma de Barcelona, E-08193, Bellaterra (Barcelona), Spain} 
\affiliation{Baylor University, Waco, Texas  76798} 
\affiliation{Istituto Nazionale di Fisica Nucleare Bologna, $^{cc}$University of Bologna, I-40127 Bologna, Italy} 
\affiliation{Brandeis University, Waltham, Massachusetts 02254} 
\affiliation{University of California, Davis, Davis, California  95616} 
\affiliation{University of California, Los Angeles, Los Angeles, California  90024} 
\affiliation{University of California, San Diego, La Jolla, California  92093} 
\affiliation{University of California, Santa Barbara, Santa Barbara, California 93106} 
\affiliation{Instituto de Fisica de Cantabria, CSIC-University of Cantabria, 39005 Santander, Spain} 
\affiliation{Carnegie Mellon University, Pittsburgh, PA  15213} 
\affiliation{Enrico Fermi Institute, University of Chicago, Chicago, Illinois 60637}
\affiliation{Comenius University, 842 48 Bratislava, Slovakia; Institute of Experimental Physics, 040 01 Kosice, Slovakia} 
\affiliation{Joint Institute for Nuclear Research, RU-141980 Dubna, Russia} 
\affiliation{Duke University, Durham, North Carolina  27708} 
\affiliation{Fermi National Accelerator Laboratory, Batavia, Illinois 60510} 
\affiliation{University of Florida, Gainesville, Florida  32611} 
\affiliation{Laboratori Nazionali di Frascati, Istituto Nazionale di Fisica Nucleare, I-00044 Frascati, Italy} 
\affiliation{University of Geneva, CH-1211 Geneva 4, Switzerland} 
\affiliation{Glasgow University, Glasgow G12 8QQ, United Kingdom} 
\affiliation{Harvard University, Cambridge, Massachusetts 02138} 
\affiliation{Division of High Energy Physics, Department of Physics, University of Helsinki and Helsinki Institute of Physics, FIN-00014, Helsinki, Finland} 
\affiliation{University of Illinois, Urbana, Illinois 61801} 
\affiliation{The Johns Hopkins University, Baltimore, Maryland 21218} 
\affiliation{Institut f\"{u}r Experimentelle Kernphysik, Karlsruhe Institute of Technology, D-76131 Karlsruhe, Germany} 
\affiliation{Center for High Energy Physics: Kyungpook National University, Daegu 702-701, Korea; Seoul National University, Seoul 151-742, Korea; Sungkyunkwan University, Suwon 440-746, Korea; Korea Institute of Science and Technology Information, Daejeon 305-806, Korea; Chonnam National University, Gwangju 500-757, Korea; Chonbuk National University, Jeonju 561-756, Korea} 
\affiliation{Ernest Orlando Lawrence Berkeley National Laboratory, Berkeley, California 94720} 
\affiliation{University of Liverpool, Liverpool L69 7ZE, United Kingdom} 
\affiliation{University College London, London WC1E 6BT, United Kingdom} 
\affiliation{Centro de Investigaciones Energeticas Medioambientales y Tecnologicas, E-28040 Madrid, Spain} 
\affiliation{Massachusetts Institute of Technology, Cambridge, Massachusetts  02139} 
\affiliation{Institute of Particle Physics: McGill University, Montr\'{e}al, Qu\'{e}bec, Canada H3A~2T8; Simon Fraser University, Burnaby, British Columbia, Canada V5A~1S6; University of Toronto, Toronto, Ontario, Canada M5S~1A7; and TRIUMF, Vancouver, British Columbia, Canada V6T~2A3} 
\affiliation{University of Michigan, Ann Arbor, Michigan 48109} 
\affiliation{Michigan State University, East Lansing, Michigan  48824}
\affiliation{Institution for Theoretical and Experimental Physics, ITEP, Moscow 117259, Russia} 
\affiliation{University of New Mexico, Albuquerque, New Mexico 87131} 
\affiliation{Northwestern University, Evanston, Illinois  60208} 
\affiliation{The Ohio State University, Columbus, Ohio  43210} 
\affiliation{Okayama University, Okayama 700-8530, Japan} 
\affiliation{Osaka City University, Osaka 588, Japan} 
\affiliation{University of Oxford, Oxford OX1 3RH, United Kingdom} 
\affiliation{Istituto Nazionale di Fisica Nucleare, Sezione di Padova-Trento, $^{dd}$University of Padova, I-35131 Padova, Italy} 
\affiliation{LPNHE, Universite Pierre et Marie Curie/IN2P3-CNRS, UMR7585, Paris, F-75252 France} 
\affiliation{University of Pennsylvania, Philadelphia, Pennsylvania 19104}
\affiliation{Istituto Nazionale di Fisica Nucleare Pisa, $^{ee}$University of Pisa, $^{ff}$University of Siena and $^{gg}$Scuola Normale Superiore, I-56127 Pisa, Italy} 
\affiliation{University of Pittsburgh, Pittsburgh, Pennsylvania 15260} 
\affiliation{Purdue University, West Lafayette, Indiana 47907} 
\affiliation{University of Rochester, Rochester, New York 14627} 
\affiliation{The Rockefeller University, New York, New York 10021} 
\affiliation{Istituto Nazionale di Fisica Nucleare, Sezione di Roma 1, $^{hh}$Sapienza Universit\`{a} di Roma, I-00185 Roma, Italy} 

\affiliation{Rutgers University, Piscataway, New Jersey 08855} 
\affiliation{Texas A\&M University, College Station, Texas 77843} 
\affiliation{Istituto Nazionale di Fisica Nucleare Trieste/Udine, I-34100 Trieste, $^{ii}$University of Trieste/Udine, I-33100 Udine, Italy} 
\affiliation{University of Tsukuba, Tsukuba, Ibaraki 305, Japan} 
\affiliation{Tufts University, Medford, Massachusetts 02155} 
\affiliation{Waseda University, Tokyo 169, Japan} 
\affiliation{Wayne State University, Detroit, Michigan  48201} 
\affiliation{University of Wisconsin, Madison, Wisconsin 53706} 
\affiliation{Yale University, New Haven, Connecticut 06520} 
\author{T.~Aaltonen}
\affiliation{Division of High Energy Physics, Department of Physics, University of Helsinki and Helsinki Institute of Physics, FIN-00014, Helsinki, Finland}
\author{J.~Adelman}
\affiliation{Enrico Fermi Institute, University of Chicago, Chicago, Illinois 60637}
\author{B.~\'{A}lvarez~Gonz\'{a}lez$^v$}
\affiliation{Instituto de Fisica de Cantabria, CSIC-University of Cantabria, 39005 Santander, Spain}
\author{S.~Amerio$^{dd}$}
\affiliation{Istituto Nazionale di Fisica Nucleare, Sezione di Padova-Trento, $^{dd}$University of Padova, I-35131 Padova, Italy} 

\author{D.~Amidei}
\affiliation{University of Michigan, Ann Arbor, Michigan 48109}
\author{A.~Anastassov}
\affiliation{Northwestern University, Evanston, Illinois  60208}
\author{A.~Annovi}
\affiliation{Laboratori Nazionali di Frascati, Istituto Nazionale di Fisica Nucleare, I-00044 Frascati, Italy}
\author{J.~Antos}
\affiliation{Comenius University, 842 48 Bratislava, Slovakia; Institute of Experimental Physics, 040 01 Kosice, Slovakia}
\author{G.~Apollinari}
\affiliation{Fermi National Accelerator Laboratory, Batavia, Illinois 60510}
\author{A.~Apresyan}
\affiliation{Purdue University, West Lafayette, Indiana 47907}
% ...

\collaboration{CDF Collaboration\footnote{With visitors from $^a$University of Massachusetts Amherst, Amherst, Massachusetts 01003,
$^b$Universiteit Antwerpen, B-2610 Antwerp, Belgium, 
$^c$University of Bristol, Bristol BS8 1TL, United Kingdom,
$^d$Chinese Academy of Sciences, Beijing 100864, China, 
$^e$Istituto Nazionale di Fisica Nucleare, Sezione di Cagliari, 09042 Monserrato (Cagliari), Italy,
$^f$University of California Irvine, Irvine, CA  92697, 
$^g$University of California Santa Cruz, Santa Cruz, CA  95064, 
$^h$Cornell University, Ithaca, NY  14853, 
$^i$University of Cyprus, Nicosia CY-1678, Cyprus, 
$^j$University College Dublin, Dublin 4, Ireland,
$^k$University of Edinburgh, Edinburgh EH9 3JZ, United Kingdom, 
$^l$University of Fukui, Fukui City, Fukui Prefecture, Japan 910-0017
$^m$Kinki University, Higashi-Osaka City, Japan 577-8502
$^n$Universidad Iberoamericana, Mexico D.F., Mexico,
$^o$University of Iowa, Iowa City, IA  52242,
$^p$Kansas State University, Manhattan, KS 66506
$^q$Queen Mary, University of London, London, E1 4NS, England,
$^r$University of Manchester, Manchester M13 9PL, England,
$^s$Muons, Inc., Batavia, IL 60510, 
$^t$Nagasaki Institute of Applied Science, Nagasaki, Japan, 
$^u$University of Notre Dame, Notre Dame, IN 46556,
$^v$University de Oviedo, E-33007 Oviedo, Spain, 
$^w$Texas Tech University, Lubbock, TX  79609, 
$^x$IFIC(CSIC-Universitat de Valencia), 56071 Valencia, Spain,
$^y$Universidad Tecnica Federico Santa Maria, 110v Valparaiso, Chile,
$^z$University of Virginia, Charlottesville, VA  22906
$^{aa}$Bergische Universit\"at Wuppertal, 42097 Wuppertal, Germany,
$^{bb}$Yarmouk University, Irbid 211-63, Jordan
$^{jj}$On leave from J.~Stefan Institute, Ljubljana, Slovenia, 
}}
\noaffiliation


\date{\today}

\begin{abstract}
We present a search for the technicolor particles $\rho_{T}$ and $\pi_{T}$ 
in the process $p\bar{p} \rightarrow \rho_{T} \rightarrow W\pi_{T}$
at a center of mass energy of $\sqrt{s}=1.96\,\mathrm{TeV}$.
The search uses a data sample corresponding to approximately 
$1.9\,\mathrm{fb}^{-1}$ of integrated luminosity accumulated by the CDF II
detector at the Fermilab Tevatron.
The event signature we consider is $W\rightarrow \ell\nu$ and
$\pi_{T} \rightarrow b\bar{b}, b\bar{c}$ or $b\bar{u}$ depending on the
$\pi_{T}$ charge.
We select events with a single high-$p_T$ electron or muon, large missing 
transverse energy, and two jets. Jets corresponding to bottom quarks are identified with multiple $b$-tagging algorithms. 
The observed number of events and the invariant mass distributions 
are consistent with the standard model background expectations, 
and we exclude a region at 95\% confidence level in the $\rho_T$-$\pi_T$ mass plane.
As a result, a large fraction of the region $m(\rho_T) = 180$ - $250\,\mathrm{GeV}/c^2$ 
and $m(\pi_T) = 95$ - $145\,\mathrm{GeV}/c^2$ is excluded.
\end{abstract}


\maketitle

The mechanism of electroweak symmetry breaking in nature
is still unknown. 
The standard model (SM) assumes the Higgs mechanism~\cite{Higgs:1964pj} but provides no
explanation as to why there should be a fundamental scalar Higgs field
with a non-zero vacuum expectation value. An alternative approach is to seek
a dynamical mechanism for the symmetry breaking.
The scenario known as technicolor~\cite{technicolor1,technicolor2,technicolor3}
proposes a new strong interaction, modeled on QCD, which spontaneously breaks electroweak
symmetry in an analogous way to the breaking of chiral symmetry in QCD.
The strong technicolor interaction between the new technifermions
results in a vacuum technifermion condensate which can break
electroweak symmetry and hence give mass to the $W^{\pm}$ and $Z$
gauge bosons.
As in QCD, the technicolor interaction should give rise to technipions
($\pi_T$) and other technimesons.
In this Letter we report the results of a search for technipions
produced in association with a $W$ boson from technirho ($\rho_T$) decay,
$\rho_T \rightarrow W \pi_T$,
in the context of the technicolor straw man (TCSM) model~\cite{tcsm}.
Like the SM Higgs boson, the technipion coupling to fermions is proportional to mass, and hence the 
technipion predominantly decays to $b\bar{b}$,
$b\bar{c}$, or $b\bar{u}$, depending on its charge.
The resulting final state is identified by selecting 
events with exactly one high-$p_T$ electron or muon candidate,
large missing transverse energy, 
and two jets, at least one of which is identified as containing a $b$-quark ($b$-tagged).

The data sample used here corresponds to $1.9 \pm0.1\,\mathrm{fb}^{-1}$  of integrated luminosity, 
nearly five times the sample used in the previous Tevatron searches~\cite{CDFTechni,D0Techni}. 
The searches at LEP were able to exclude $\rho_T$ 
production at 95\% confidence level for $90<m_{\rho_T}<206.7\,\mathrm{GeV}/c^2$, 
independently of the assumed $\pi_T$ mass and other model parameters~\cite{DelphiTechni}. 

The CDF II~\cite{CDFDet} is a general purpose detector 
to study $p\bar p$ collisions at $\sqrt{s}=1.96\,\mathrm{TeV}$ at the Fermilab Tevatron.
It consists of a cylindrical magnetic spectrometer, 
surrounded by electromagnetic and hadronic calorimeters.
Charged particle tracking is performed with microstrip silicon detectors surrounded by a large
cylindrical multilayer drift chamber, both immersed in a 1.4 T solenoidal magnetic field 
aligned coaxially with the incoming beams.
Jets are identified as collections of electromagnetic and hadronic energy deposits in calorimeter towers,
which are clustered using an iterative cone algorithm with a cone of 
$\Delta R = \sqrt{(\Delta\phi)^2 + (\Delta\eta)^2} = 0.4$~\cite{coordinates}.
Muons are identified by a system of drift chambers and scintillators placed outside 
the calorimeter 
at a depth of at least five nuclear interaction lengths from the interaction region.

Events are collected using high-$p_T$ electron or muon triggers with a three-level selection filter.
The first two levels identify purely electromagnetic calorimeter clusters, or require that 
track segments in the muon chambers align with tracks in the drift chamber having $p_T>8\,\mathrm{GeV}/c$.
The third-level trigger requires an electron (muon) with $E_T > 18\,\mathrm{GeV}$ ($p_T > 18\,\mathrm{GeV}/c$).

Events are further required to have exactly one electron or muon candidate, 
large missing transverse energy ($\met > 20\, \mathrm{GeV}$)~\cite{coordinates}, 
and two jets.
The electron or muon must be within the central part of the detector, in the pseudorapidity regions $|\eta|<1.1$
or $|\eta|<1.0$, respectively, and must have $E_T >20\, \mathrm{GeV}$ or $p_T >20\,\mathrm{GeV}$. 
Because the lepton from a leptonic $W$ decay is  well isolated from the rest of the event, the energy deposits in 
calorimeter towers within the cone of 
$\Delta R = 0.4$ surrounding the lepton is required to contain less than 10\% of the lepton energy.
It must also be no more than 5 cm in $z$ away from the primary event vertex, which is 
defined by fitting a subset of charged particle tracks in the event to a single vertex. 
To reduce the background from $Z$ boson decays, we reject not only events 
with multiple high-$p_T$ leptons, 
but also events in which the lepton and another high-$p_T$ track 
of opposite sign form an invariant mass between $76 < M_{ll} < 106\, \mathrm{GeV}/c^2$. 
Jets used in the analysis must fall within the acceptance of the silicon detector ($|\eta|<2.0$) 
for reliable $b$-tagging, and they must have transverse energy greater than
$20\,\mathrm{GeV}$.

The primary background to this technicolor search is SM $W$ + 2~jets production. However this process is dominated by 
light-flavor jets,
while the technipion decay process should contain at least one $b$ quark. 
Identifying these $b$-quark jets therefore helps to significantly suppress the background. 
We use two $b$-tagging algorithms: a secondary vertex finding algorithm~\cite{Higgs1.0fb}
({\sc secvtx}) and a jet probability tagging algorithm~\cite{Affolder:2001wd} ({\sc jetprob}).
To further improve the purity of the {\sc secvtx} sample,
a neural network (NN) filter has been 
trained to reject tagged jets originating from charm or light-quarks~\cite{Higgs1.0fb}.
The search sensitivity is maximized by using three exclusive $b$-tagged event categories.   
The first category (ST+ST) contains events with two {\sc secvtx} $b$-tagged jets. 
The second category (ST+JP) consists of events where only one of the jets 
is $b$-tagged by {\sc secvtx} and the second jet is only $b$-tagged by {\sc jetprob}. 
The third category (ST+NNtag) is for events which do not belong to the first two categories but 
contain exactly one {\sc secvtx} $b$-tagged jet
that also passes the NN filter. 

The selected event sample includes contributions from other standard model processes. 
The largest backgrounds are due to 
$W$ + jets production, $t\bar t$ production, and non-$W$ multijet production, with small contributions from 
single top, diboson ($WW$, $WZ$, $ZZ$), and $Z\rightarrow \tau\tau$ production. 
These backgrounds are estimated using the same methods as  
the standard model Higgs boson search analysis in the $W$ + 2~jets channel~\cite{Higgs1.9fb}.
A summary of the estimated backgrounds to the $W$ +
2~jets final state is shown in Table~\ref{tbl:background}, along
with the number of observed events in data and the expected technicolor signal events for 
a mass hypothesis of  $m(\rho_T,\pi_T)=(180,95), (200,115), (250,145)\,  \mathrm{GeV}/c^2$, respectively.

\begin{table*}
  \begin{center}
\begin{tabular}{cccc}\hline \hline
          Selection &        ST+ST        &   ST+JP       & ST+NNtag \\ \hline
         $Wb\bar{b}$&      37.9$\pm$16.9& 31.2$\pm$14.0&  215.6$\pm$92.3 \\
         $Wc\bar{c}$&        2.9$\pm$1.2& 7.9$\pm$3.4&  167.0$\pm$62.1\\
              Mistag&        3.9$\pm$0.4& 11.7$\pm$0.9&  107.1$\pm$9.4\\
   $t\bar{t}$&       19.0$\pm$2.9& 15.6$\pm$2.4&  60.7$\pm$9.3\\
   Single top &               8.5$\pm$1.2& 7.0$\pm$1.0&  44.0$\pm$6.4 \\
            non-W &        5.5$\pm$1.0& 9.6$\pm$1.7&  184.7$\pm$33.0   \\ 
                  WW&        0.17$\pm$0.02& 0.9$\pm$0.1&  15.4$\pm$1.9\\
                  WZ&        2.41$\pm$0.26& 1.8$\pm$0.2&   7.6$\pm$0.8\\
                  ZZ&        0.06$\pm$0.01& 0.08$\pm$0.01&  0.31$\pm$0.03 \\
       $Z\rightarrow \tau\tau$&        0.25$\pm$0.04& 1.3$\pm$0.2&  7.3$\pm$1.1\\ \hline
           Total Bkg. &      80.6$\pm$18.8& 87.0$\pm$18.0&  809.6$\pm$159.4 \\ \hline
$m({\rho_T},\pi_T)=(180,95)$ &  22.9$\pm$2.9 & 19.5$\pm$2.6 & 81.9 $\pm$ 5.6 \\
$m({\rho_T},\pi_T)=(200,115)$ &  12.7$\pm$1.6 & 10.5$\pm$1.4 & 43.7 $\pm$ 3.0 \\
$m({\rho_T},\pi_T)=(250,145)$ &  7.7$\pm$1.0 & 6.6$\pm$0.9 & 27.3 $\pm$ 1.9 \\ \hline

     Data Events&      83&              90&                 805 \\ \hline\hline
\end{tabular}
  \caption{Predicted sample composition and observed number of $W$ + 2~jets in each $b$-tagging category, along with the expected 
 signal events for a mass hypothesis of $m(\rho_T,\pi_T)=(180,95), (200,115), (250,145)\,  \mathrm{GeV}/c^2$, respectively.}
  \label{tbl:background}
  \end{center}
\end{table*}

\begin{figure*}[htbp]
 \begin{center}
  \includegraphics[width=0.40\textwidth]{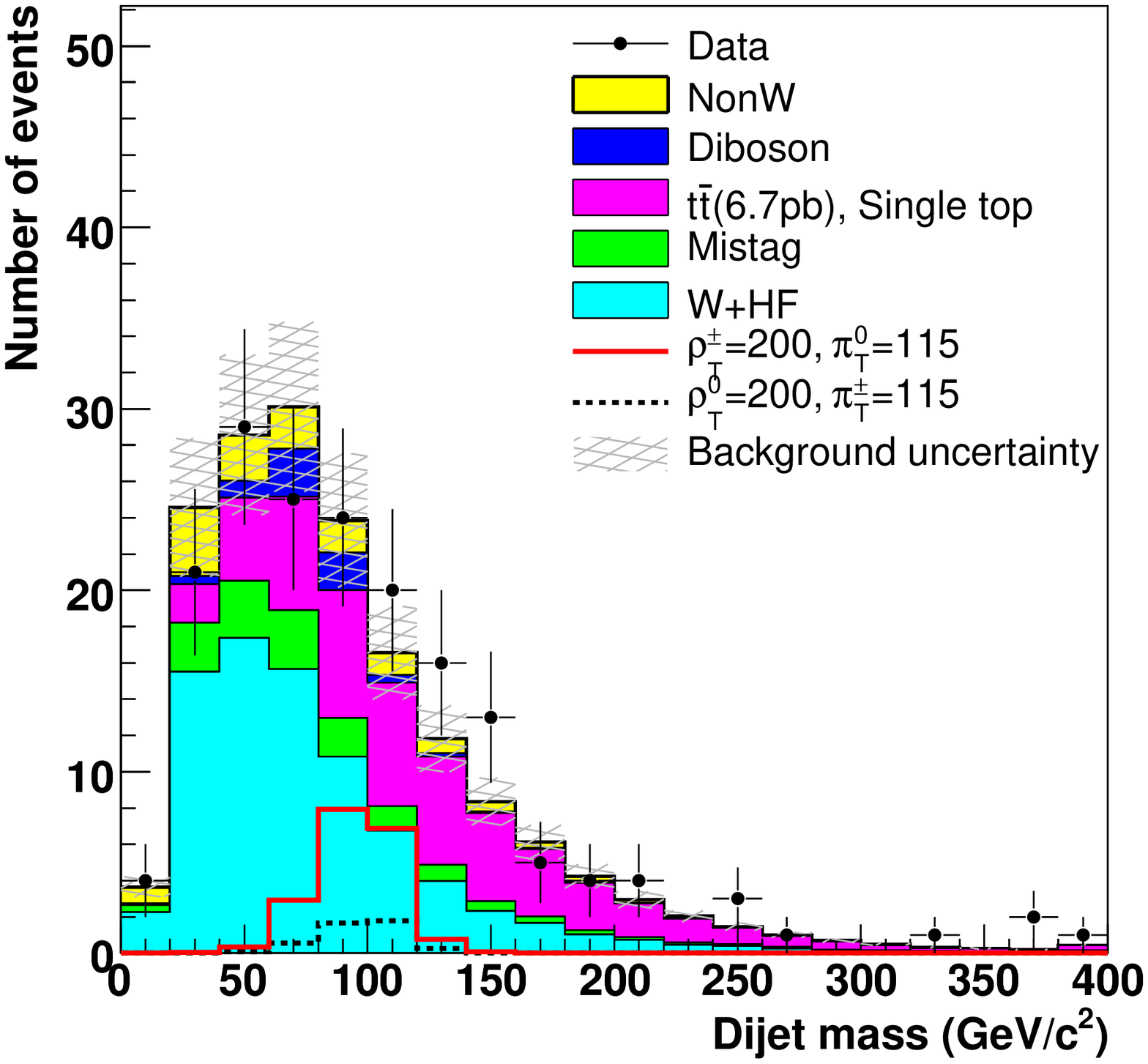}
  \includegraphics[width=0.40\textwidth]{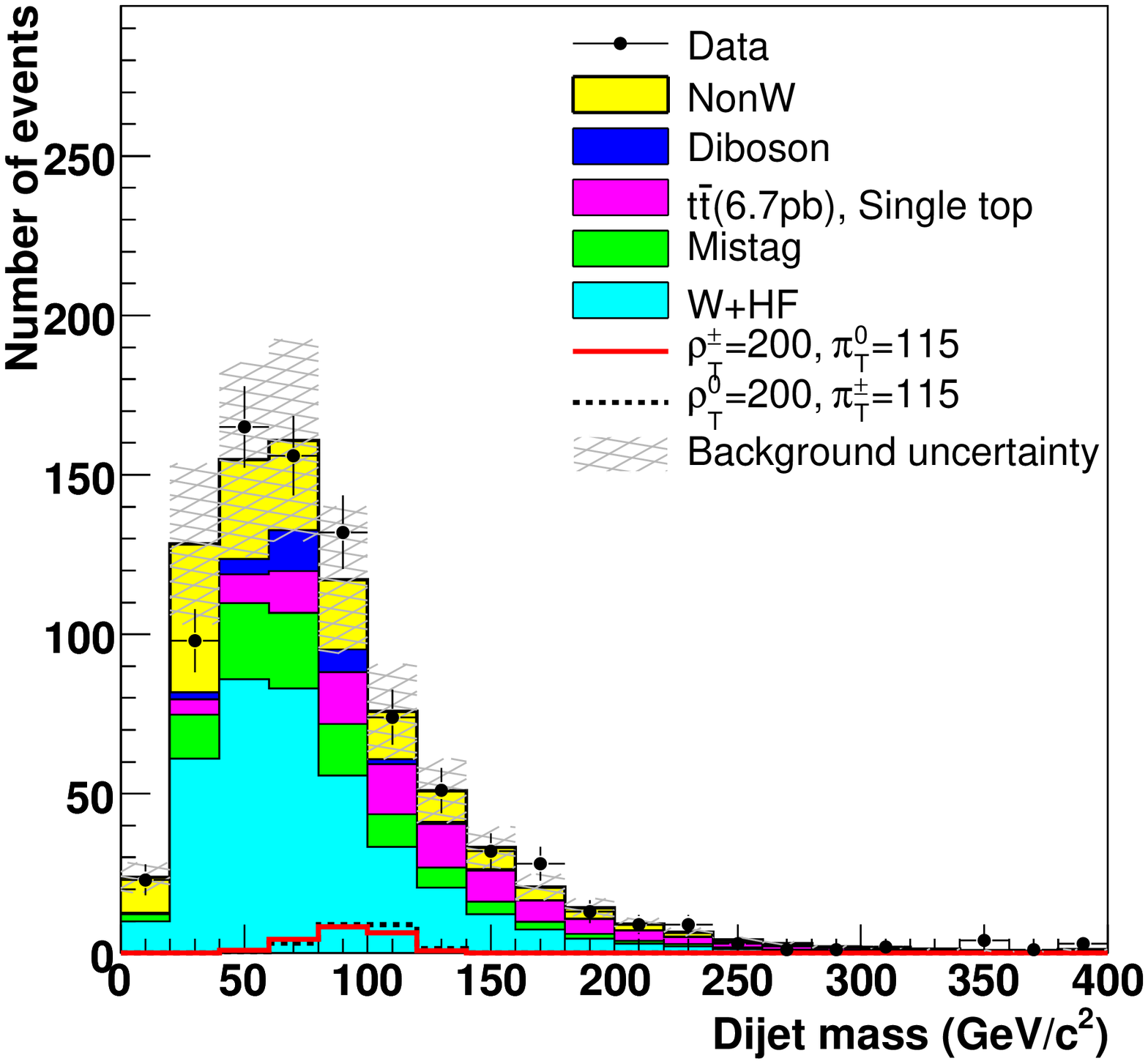}
  \caption{Reconstructed dijet mass distributions for $W$ + 2~jets events.
 The left is for double tags (ST+ST and ST+JP) and the right is for single tag (ST+NNtag) events.}
  \label{fig:dijetmass} \end{center}
\end{figure*}

\begin{figure*}[htbp]
 \begin{center}
  \includegraphics[width=0.40\textwidth]{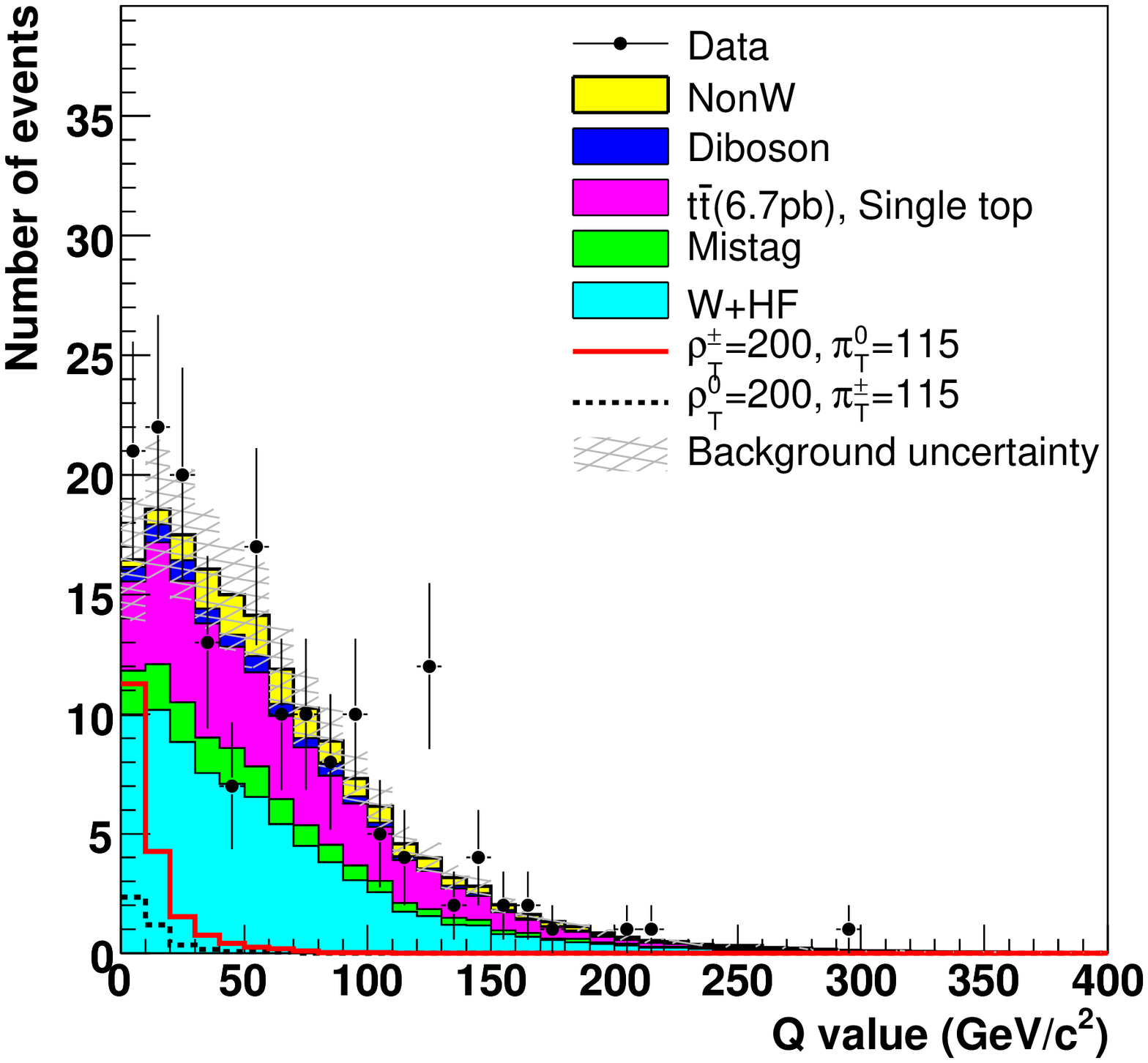}
  \includegraphics[width=0.40\textwidth]{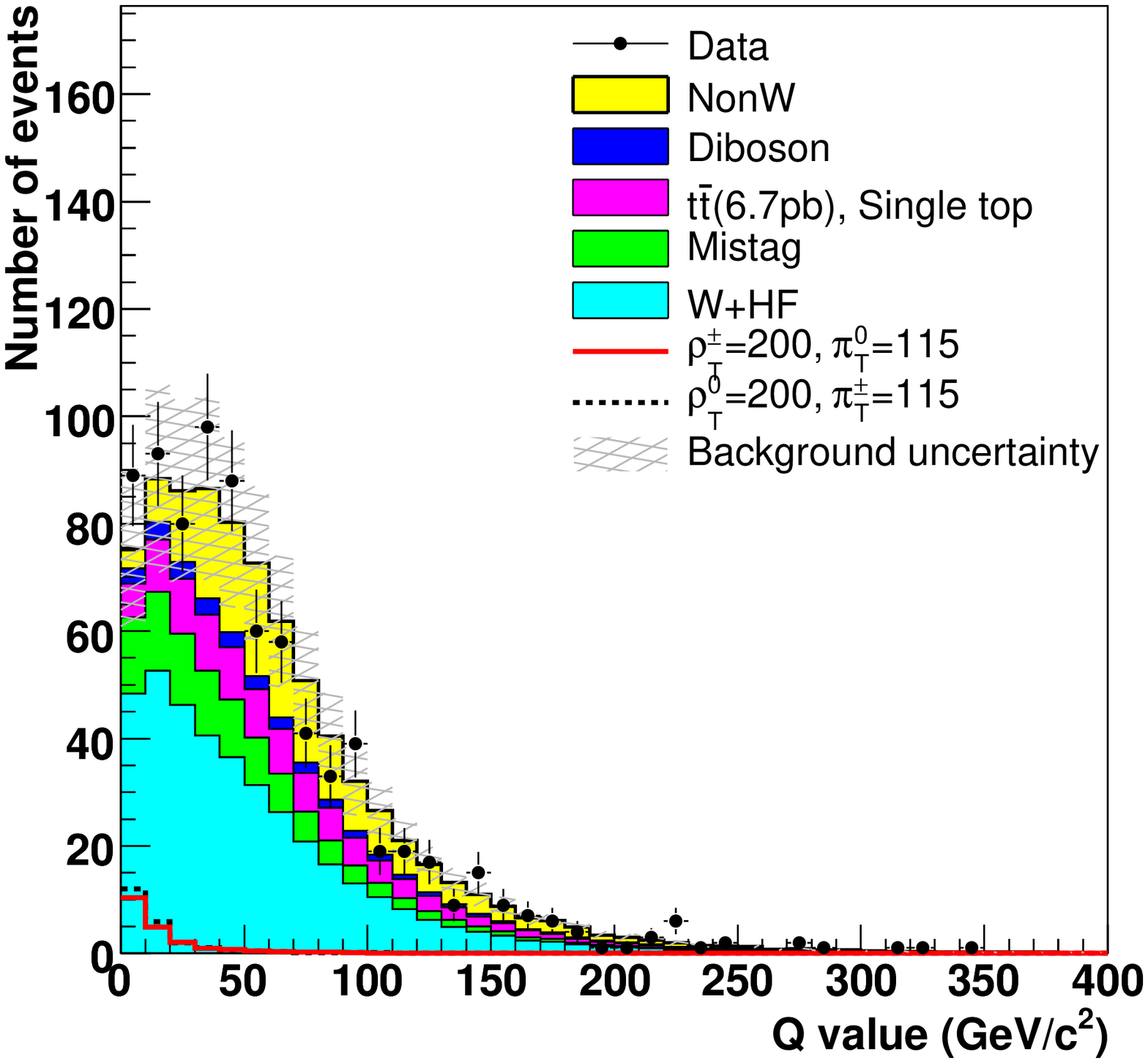}
  \caption{Reconstructed $Q$-value distributions for $W$ + 2~jets events, 
  where $Q = m(\rho_T) - m(\pi_T) - m(W)$.
 The left is for double tags (ST+ST and ST+JP) and the right is for  
 single tag (ST+NNtag) events.}
  \label{fig:Qval} \end{center}
\end{figure*}

The signal process in $\rho_T \rightarrow W \pi_T\rightarrow l\nu j_1 j_2$ is expected to show 
resonant peaks in both the dijet and $W$ + 2~jets mass spectra.  We reconstruct the $p_z$ of the 
neutrino by constraining the invariant mass of the lepton-neutrino pair to the $W$ boson mass, which 
gives a two-fold ambiguity. We select the solution 
with the smaller $|p_z|$, since that is more probable given the production mechanism 
of this heavy state; if there is no real solution, we set the imaginary part of solution to zero. 
Figure~\ref{fig:dijetmass} shows the observed dijet mass
spectra in the double tagged (ST+ST and ST+JP) and one {\sc secvtx} with NN filter tagged  
2 jets samples, along with the distributions expected from the background processes.  
Figure~\ref{fig:Qval} shows the $Q$-value distribution in each $b$-tagging category, 
the mass difference defined as $Q = m(\rho_T) - m(\pi_T) - m(W)$, which exploits the fact that the 
$Q$-value for the $\rho_T$ decay is quite small and consequently the resolution of the mass difference
is better than the mass of the $\rho_T$ itself. 
The signal distributions from the charged and neutral technicolor particles with 
$m_{\rho_T} = 200\,\mathrm{GeV}/c^2$ and $m_{\pi_T} = 115\,\mathrm{GeV}/c^2$ are also shown for comparison.
There is no significant excess observed in either the dijet mass or $Q$-value distributions.


The acceptance for $\rho_{T} \rightarrow W\pi_{T} \rightarrow \ell \nu b\bar b, b\bar c, b\bar u$
is calculated from samples generated with the {\sc pythia} Monte Carlo program~\cite{Sjostrand:2000wi}
using  $\rho_{T}$ mass values between $180$ and $250\,\mathrm{GeV}/c^2$ with a step of 10 $\mathrm{GeV}/c^2$, and 
 $max(m(\rho_T)/2, m(W))<m(\pi_T)<m(\rho_T)-m(W)$ where the decay $\rho_T\rightarrow W\pi_T$ dominates.

The total acceptances for ST+ST, ST+JP 
and ST+NNtag events of $\pi_{T}^{0} \rightarrow b\bar b$ ($\pi_{T}^{\pm} \rightarrow b\bar c, b\bar u$) are 
$0.32-0.45\%\, (0.04-0.06\%)$, $0.23-0.31\%\, (0.09-0.13\%)$, and $0.66-0.81\%\, (0.69-0.83\%)$, increasing
linearly as function of $m(\rho_T, \pi_T)$ from (180,95) to (250,165) $\mathrm{GeV}/c^2$.
The dominant systematic uncertainty on the acceptance for the 
$\pi_{T}^{0} \rightarrow b\bar b$ ($\pi_{T}^{\pm} \rightarrow b\bar c, b\bar u$) process 
originates from the uncertainty on the $b$-tagging efficiencies, 
which is a 8.4\% (9.4\%) relative error for ST+ST,
a 9.2\% (17.0\%) relative error for ST+JP, 
and a 4.3\% (4.3\%) relative error for ST+NNtag.
Additional sources of systematic error
include the jet energy scale, the lepton identification efficiency, parton distribution functions,
and the initial and final state radiation
models~\cite{Bhatti:2005ai}. The systematic uncertainties associated with the shape of dijet invariant mass and $Q$-value are also 
studied by varying the jet energy scale and the initial and final state radiation, which are  
found to have a negligible impact on the final results. 

Since there is no significant excess of events in the data compared to the predicted background, 
we set the 95\% C.L. excluded region on technicolor production 
as a function of the technicolor particle mass.
A 2-dimensional binned maximum-likelihood technique which assumes Poisson statistics is used 
on the 2-dimensional distribution of dijet invariant mass vs $Q$-value
by constraining the number of background events within the uncertainties. 
To calculate the 95\% C.L. excluded region, we use neutral and charged $\pi_T$ signals simultaneously.
A Bayesian interval is constructed
from the cumulative likelihood distributions and a prior probability
density function uniform in the number of technicolor signal events.
The 95\% confidence level upper limit is defined to be the value
$s_\mathrm{up}$ for which $\int_0^{s_\mathrm{up}} L(s) ds /
\int_0^\infty L(s) ds = 0.95$.  
The number of signal events is then converted to a technicolor particle
production cross section times branching fraction $\sigma(p\bar
p\rightarrow W\pi_T^{0}(\pi_T^{\pm}) ) \cdot BR(\pi_T^{0}(\pi_T^{\pm})\rightarrow b\bar b (b\bar c, b\bar u))$.

\begin{figure}[htbp]
 \begin{center}
  \includegraphics[width=0.45\textwidth]{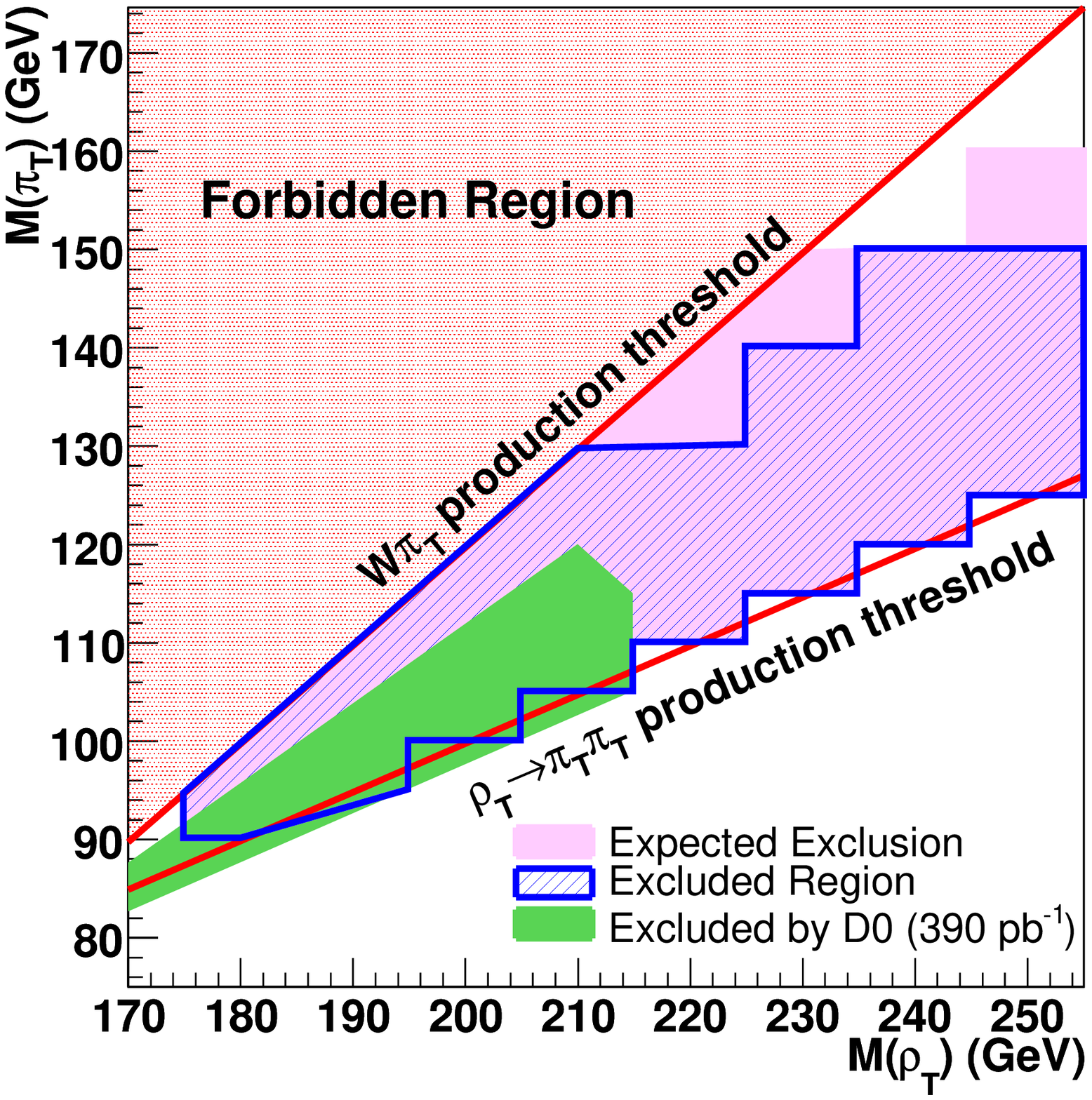}
  \caption{95\% confidence level excluded region on technicolor particles production
  cross section times branching fraction as a function of $m(\rho_T)$ and $m(\pi_T)$
  mass hypothesis.  The expected excluded region from background-only
  pseudoexperiments are shown with the observed results from
  this analysis and D0 searches.}  \label{fig:limit}
  \end{center}
\end{figure}

The expected and observed 95\% confidence level excluded region in the $\rho_T$-$\pi_T$ mass plane
is shown in Fig.~\ref{fig:limit}.
Almost the entire region we have looked at in this search
is excluded at 95\% confidence level,
except the area near the $W \pi_T$ production threshold 
with $m(\rho_T) \geq 220\,\mathrm{GeV}/c^2$ and $m(\pi_T) \geq 125\,\mathrm{GeV}/c^2$.

In summary, we have performed a search for technicolor production
$p\bar p \rightarrow \rho_T^{\pm/0} \rightarrow W^\pm \pi_T^{0/\mp} \rightarrow \ell \nu
b\bar b$, $\ell \nu b\bar c$, or $\ell \nu b\bar u$ 
using $1.9\,\mathrm{fb}^{-1}$ of integrated luminosity accumulated by the CDF II detector.
A large fraction of the region of $m(\rho_T) = 180$ - $250\,\mathrm{GeV}/c^2$ and $m(\pi_T) = 95$ - $145\,\mathrm{GeV}/c^2$ 
is excluded at 95\% confidence level, based on the technicolor Straw Man model. 
This measurement excludes a much larger region than the previous Tevatron searches~\cite{CDFTechni,D0Techni}. 

\begin{acknowledgments}
We thank Ken Lane for many fruitful discussions, 
the Fermilab staff and the technical staffs of the participating institutions for their vital 
contributions. This work was supported by the U.S. Department of Energy and National Science Foundation; the Italian 
Istituto Nazionale di Fisica Nucleare; the Ministry of Education, Culture, Sports, Science and Technology of Japan; 
the Natural Sciences and Engineering Research Council of Canada; the National Science Council of the Republic of 
China; the Swiss National Science Foundation; the A.P. Sloan Foundation; the Bundesministerium f\"ur Bildung und 
Forschung, Germany; the World Class University Program, the National Research Foundation of Korea; the Science and 
Technology Facilities Council and the Royal Society, UK; the Institut National de Physique Nucleaire et Physique 
des Particules/CNRS; the Russian Foundation for Basic Research; the Ministerio de Ciencia e Innovaci\'{o}n, and 
Programa Consolider-Ingenio 2010, Spain; the Slovak R\&D Agency; and the Academy of Finland. 
\end{acknowledgments}


\begin{thebibliography}{20}
\bibitem[{coo()}]{coordinates}
\bibinfo{note}{The CDF reference frame uses cylindrical coordinates, where $z$
  is the direction of the proton beam, and $\theta$ and $\phi$ are the polar
  and azimuthal angles with respect to the proton beam. The pseudorapidity is
  defined as $\eta = -\ln \tan \theta/2$, the transverse energy $E_T=E\sin
  \theta$, and the transverse momentum $p_T=p \sin \theta$. 
  The missing transverse energy magnitude (\met) 
  is the vector sum of all calorimeter deposits, projected into the transverse
  plane. The \met ~vector is corrected for the energy deposition of
  high-$p_T$ muons as well as for jet energy corrections.}
\end{thebibliography}

\end{document}